\documentclass[smallextended,final]{svjour3}
\usepackage[utf8]{inputenc}
\usepackage[greek,english]{babel}
\usepackage{graphicx}
\usepackage{textcomp}
\usepackage{color}
\usepackage{amssymb,amsmath}
\usepackage{hyperref}
\usepackage{times}
\usepackage[authoryear,round]{natbib}
\usepackage{multirow}
\usepackage{arydshln}
\usepackage{enumitem}
\title{A simple model of scientific progress -- with examples\footnote{The present article is an extended version
    of \citep{Scorzato-silfs14}, as it includes, in particular, a variety of significant examples.}}
\author{Luigi Scorzato}
\institute{Present address: Accenture AG, Digital, Big Data Analytics. 20 Rue de Pr{\'e}-Bois CH-1215 Gen{\`e}ve.
\email{luigi@scorzato.it}} 
\date{}

\begin{document}

\newtheorem{defn}{Definition}
\newtheorem{post}{Postulate}
\newtheorem{rmrk}{Remark}

\maketitle

\begin{abstract}
One of the main goals of scientific research is to provide a description of the empirical data which is as accurate
and comprehensive as possible, while relying on as few and simple assumptions as possible.  In this paper, I
propose a definition of the notion of {\em few and simple assumptions} that is not affected by known problems.
This leads to the introduction of a simple model of scientific progress that is based only on empirical accuracy
and conciseness.  An essential point in this task is the understanding of the role played by {\em measurability} in
the formulation of a scientific theory.  This is the key to prevent artificially concise formulations.  The model
is confronted here with many possible objections and with challenging cases of real progress.  Although I cannot
exclude that the model might have some limitations, it includes all the cases of genuine progress examined here,
and no spurious one.  In this model, I stress the role of the {\em state of the art}, which is the collection of
all the theories that are the only legitimate source of scientific predictions.  Progress is a global upgrade of
the state of the art.
\end{abstract}

\section{Introduction}

What makes science very special, among human activities, is the possibility of being progressive.  This does not
mean that scientific progress comes naturally: sometimes precious knowledge is lost\footnote{See \citet{RussoFR},
  for an impressive reconstruction of the loss of the Hellenistic science.}; sometimes, experimental or theoretical
results are revised.  However, these steps back do not refute the progressive nature of science: they rather urge
even more the quest for understanding it.

However, a characterization of scientific progress has proven extremely elusive, since \citet{Kuhn1} convincingly
showed that the naive idea of progress as an {\em accumulation of empirically verified statements} is
untenable\footnote{The idea that scientific progress consists in accumulation of knowledge has been revived a few
  years ago \citep{BirdNous}, but the problem remains to understand {\em which} knowledge is relevant, since
  non-justified knowledge should not count \citep{Bird2008-BIRSPA}, and it is not clear what are justified beliefs,
  without relying, in turn, on a notion of knowledge \citep{Bird2007-BIRJJ}.  See also
  \citet{Niiniluoto2014,Rowbottom2015}, for a recent update on this debate.}.  For example, even the very
successful Newton's theory of gravity needed to be restricted to velocities much smaller than the speed of light,
in order to retain validity.  Although we believe that this restriction is not a defeat --- quite the contrary ---
it certainly precludes a characterization of progress in terms of accumulation of valid sentences.

Much progress is being achieved by deriving new scientific results from a limited set of laws that are regarded as
fixed and fundamental.  But, while this {\em reductionist} \citep{reductionism-IEP} model of progress may describe
a substantial part of what \citet{Kuhn1} called {\em normal science}, it completely fails to characterize many of
the most remarkable achievements that are unanimously regarded as progressive.  This is the case, in particular,
when new fundamental laws are discovered that supersede what was previously considered fundamental, or when new
laws are found that are empirically very successful, although we are unable to reduce them to more fundamental
ones.

An appealing view suggests that there is scientific progress when new theories are discovered that are better
than the available ones \citep{PopperCR}.  But, what does {\em better} mean?  It is certainly not enough to
characterize better theories in terms of empirical accuracy.  In fact, if we take seriously the idea that only the
agreement with the experiments matters, to evaluate scientific theories, then the {\em bare collection of all the
  available experimental reports} should always be the best possible ``theory''\footnote{Experimental reports never
  contradict each others, as long as they bear different dates or locations.}.  However, we certainly do not regard
such unstructured collection as a valuable theory, if for no other reason, because it enables no prediction.

This suggests to characterize better theories as those that lead to {\em novel predictions}.  The idea is very
appealing, and it was at the heart of Lakatos' view of progressive research programmes \citep{Lakatos_1970}.  There
is little doubt that successful predictions are exactly what scientists regard as the greatest reward.  But, how do
we use successful predictions for theory selection?  Nobody ever formulated a convincing proposal in this
sense\footnote{As far as I know, the most serious attempt in this direction is the discussion of {\em dynamic
    consilience} in \citep{Thagard1978-THATBE}.  However, Thagard admits that a precise definition is very
  difficult, and no argument is given to justify why the following example should not be considered a successful
  prediction.}, and for a good reason: it is too easy to generate formal predictions by brute force.  To illustrate
this, consider an already empirically accurate theory (that can always be produced by patching together various
models, each with limited applicability, and by adding ad-hoc assumptions to account for any remaining anomalies),
and imagine that there will be a new experiment soon, for which that theory makes no prediction.  A professor could
assign to each one of his many students, the extension of that theory in many possible ways, by introducing many
different ad-hoc assumptions.  In this way, these students might even cover the whole range of possible outcomes of
the upcoming experiment.  At least one of such ``theories'' will hit the result that will be eventually measured!
Is this a valuable prediction?!  Obviously not, but, how can we tell the lucky student who got it right that it was
only by chance?  Similarly, how can we tell the clairvoyant who predicted the earthquake that, considering all the
failed predictions by {\em other} clairvoyants, his success is meaningless?  {\em His own} record of successes is
100\%.  Also Einstein's General Relativity (GR) made only one {\em impressive} prediction\footnote{The bending of
  light in 1919.  After that, it is easy to build other theories that share the same predictions as GR.}.  What
makes GR vastly more valuable than the many lucky guesses we hear about every day is not the astonishment of the
prediction itself: it is a {\em valuable theory} behind.

In fact, science is not defenseless against those brute force attacks and clairvoyants' claims.  We are usually
able to spot them easily, but to do that, we need to rely on more than old or new empirical evidence: we need to
rely --- wittingly or not --- on some {\em cognitive values} that are {\em non-empirical}
\citep{putnam2002collapse}.  But which ones?  Failure to recognize these values represents a big threat on the
reliability of the scientists' claims, and the scientific process overall.  Unfortunately, there seems to be no
agreement on what should count here.  The cognitive values of {\em simplicity of the
  assumptions}\footnote{Sometimes, reference to {\em simplicity} is made to justify propositions that are actually
  justifiable in terms of likelihood \citep{Sober1990-SOBEIB}.  But, as noted above, likelihood is never enough to
  justify theory selection.  Here, non-empirical cognitive values always refer to values that {\em cannot} be
  justified in terms of likelihood.}, their {\em parsimony}, {\em elegance}, {\em explanatory power}, etc. are
often emphasized \citep{simplicity-IEP,Baker-SEP,Sober_PoS,GoodmanSimple}, but there is no agreement on what these
concepts mean\footnote{See, e.g., \citep{Thagard1978-THATBE}, where {\em simplicity} is recognized as the most
  important constraint, but its characterization is not sufficiently precise to tell why a long collection of
  experimental reports cannot be regarded as simple as any respectable scientific theory, on a suitable metric.}.
Quite enigmatically, \citet{KuhnET} stated that these values are necessarily imprecise.  But what does {\em
  imprecise} mean? The word {\em 'imprecise'} differs from {\em 'totally arbitrary'} only because the former
necessarily assumes a limited range of possibilities (at least in a probabilistic sense).  If that were the case,
we could certainly exploit that limited range to justify many cases of theory selection and define scientific
progress!  But, unfortunately, nobody ever defined that range.  On the contrary, according to well known general
arguments (that we will review below; see also \citealp{Kelly-razor}), for any theory $T$, and for a wide class of
notions of complexity it is always possible to chose a language in which the complexity of $T$ becomes trivially
small.  Hence, where Kuhn writes {\em imprecise}, we are apparently forced to read {\em totally arbitrary}.
Indeed, if we cannot restrict the notions of complexity somehow, the resulting arbitrariness in the cognitive
values leads inevitably to almost arbitrary theory selection\footnote{The status of the value of {\em explanatory
    power} is not better, since it also needs some notion of simplicity to be defined.  See e.g. the notion of {\em
    lovelier} in \citep{lipton2004inference}, and the discussion in Sec.~\ref{sec:ogos}.  See also
  \citet{CrupiTentori} and \citet{SchupSpreng}}.

To illustrate better this key point, consider the example of the Standard Model of particle physics (SM), which can
be defined in terms of a rather lengthy Lagrangian (see, e.g., \citealp{cottingham2007introduction}).  The SM
represents a spectacular unification of a huge variety of phenomena and it currently agrees, with remarkable
precision, with all the experiments.

The problems of the SM are indeed {\em non-empirical}.  They are the lack of an elegant unification with General
Relativity\footnote{The SM is not necessarily in {\em contradiction} with (classical) General Relativity: a
  patchwork theory made of both these theories (combined with any prescription resolving the ambiguities that
  emerge in contexts that are anyway not yet experimentally accessible) is ugly, cumbersome, but neither
  contradictory nor in conflict with any experiment.}, the lack of {\em naturalness}\footnote{Naturalness is not
  precisely defined.  Two possibilities are discussed by \citet{Bardeen:1995kv}.}, and the presence of about thirty
free parameters\footnote{Some curiously vanishing; some curiously much much bigger than others.}  that must be
determined from the experiments.  Since none of these is a problem of empirical accuracy, it is essential to
understand what are the {\em non-empirical cognitive values} associated to them.  It is difficult to answer this
question, because, in principle, we could solve all these problems by rewriting our fundamental laws as $\Xi=0$,
where all the fundamental equations of the SM and General Relativity are mapped into a single variable $\Xi$.  In
fact, nothing prevents us to define $\Xi$ as a whole set of complex equations\footnote{If I accept that the SM
  equations $Eq_i$ ($i=1,\ldots,n$) are necessarily known with finite precision $10^{-p}$, I can even represent all
  of them with a single real number: first set to zero any digit beyond the $p$-th in each equation, then set
  $\Xi=Eq_1 + Eq_2 10^{-p} + Eq_3 10^{-2p} +...$}.  Superficially, this is the most elegant and parameter-free
formulation we can think of! One could object that $\Xi$ is not directly measurable, and that we can only translate
it back to the traditional notation, to assign a value to it.  But, what does {\em directly measurable} mean?  The
translation of $\Xi$ is certainly not more complex than building the Large Hadron Collider and interpreting its
collision events.  Shall we call {\em directly measurable} only what is seen by a human being without instruments?
Isn't the human eye already an instrument (and a very sophisticated one indeed)?  We must clarify these issues, if
we want to show, convincingly, that the goal of improving the SM is not dictated by subjective taste.

These conclusions are by no means restricted to the SM or to particle physics.  In fact, for any empirical
scientific discipline, we can always produce an empirically impeccable theory by patching partial models together
and resorting to ad-hoc assumptions and lists of exceptions to save any remaining anomalies.  In a suitable
$\Xi$-like formulation, that patchwork theory becomes both the most accurate and the simplest possible
theory\footnote{This argument is sometimes used to argue for a {\em semantic} notion of simplicity, rather than a
  {\em syntactic} one.  But unfortunately, nobody ever defined precisely a semantic notion that is able to escape
  this problem.  See also \citep{Lutz-ss} about semantic vs. syntactic views.}.  What do we need to improve?  I
stress that $\Xi$-like formulations do represent a problem only for some special view of progress, but for any
meaningful cognitive value and hence for any serious attempt to characterize the goals of science.  Indeed, if
someone tells us that he can interpret $\Xi$ (and we have no clear reason to refute that), and observes that the
law $\Xi=0$ is accurate, how could {\em any} cognitive value be less than optimal in such theory?  On which grounds
could we deny that this is the ultimate best theory of all?

In front of these difficulties it is tempting to resort to {\em vague} cognitive values, that we avoid to define
precisely.  But this is acceptable only as long as we can still hope that these values could remain meaningful
under a closer inspection.  But, if we already know --- as it is the case here --- that a closer inspection can
only confirm the same problems discussed above, proposing values that lack a precise definition is like ignoring
the elephant in the room.

Clearly, we feel that the simplification brought about by $\Xi$ is artificial, and that the idea of simplicity ---
in spite of its ambiguities --- is {\em not totally arbitrary}.  Can we make this feeling precise?  What is wrong
with $\Xi$?  Does it have an {\em intrinsic} problem that limits its usability or its value? And, if so, which one?
Or is it just {\em our subjective taste} (biased by our cultural legacies) that prevents us to appreciate that odd
language?  And, if so, why not getting used to $\Xi$? How difficult would it be to build a $\Xi$-meter?

\begin{figure}
\centering
\includegraphics[width=100mm]{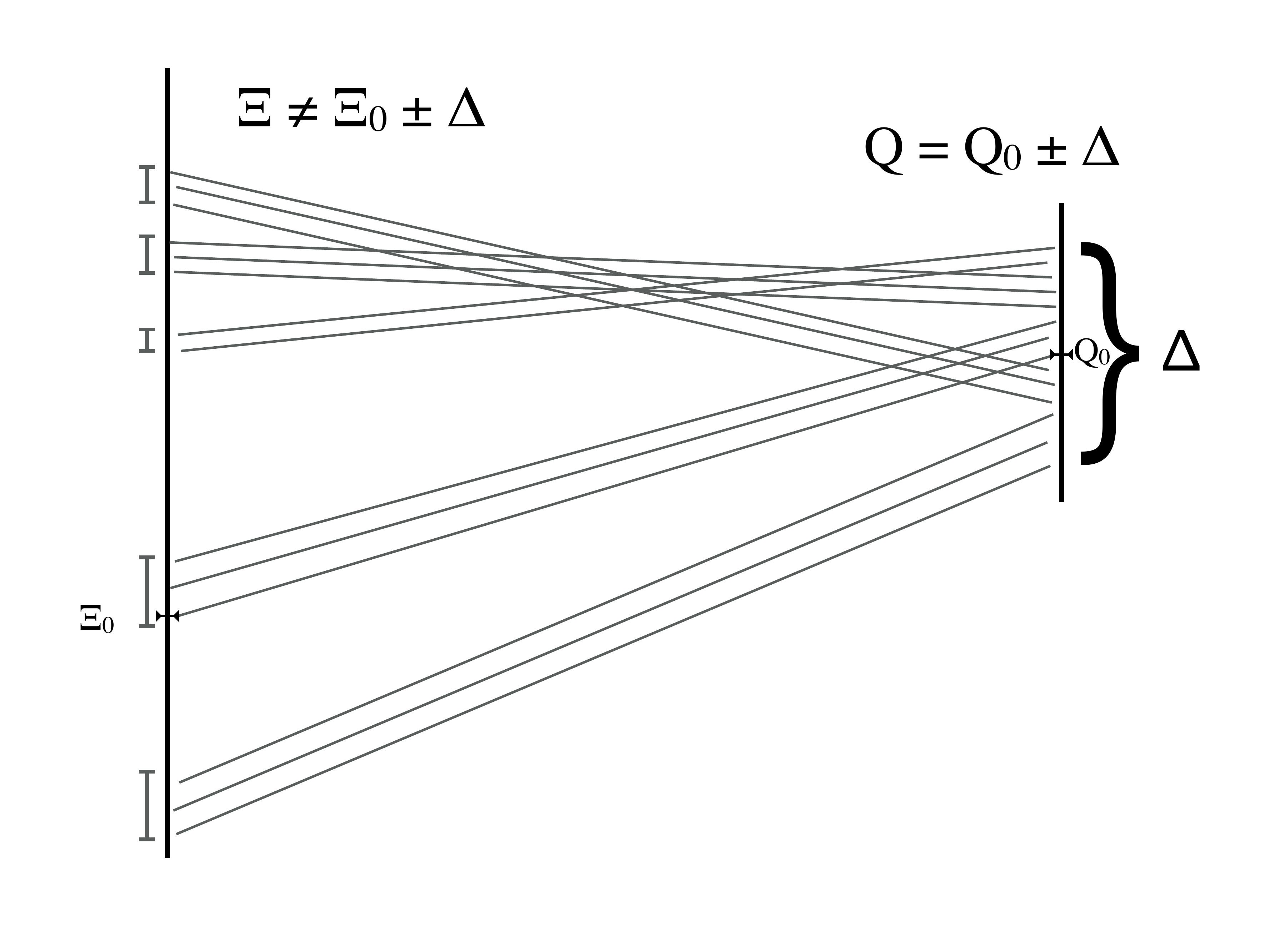}
\caption{The translation between the formulation in terms of $Q$ and the one in terms of $\Xi$ associates a value
  $\Xi_0$ to a value $Q_0$.  But, a connected interval in $Q$ is not necessarily mapped into a connected interval
  in $\Xi$.}
\label{fig:disc}
\end{figure}

As a matter of fact, interpreting $\Xi$ and building a $\Xi$-meter is not only difficult, but {\em impossible} for
a very fundamental reason \citep{Scorzato}: in general, its measurements cannot be reported in the
form\footnote{I.e., they cannot be written as a central value $\Xi_0$ and a {\em connected} errorbar of some size
  $\Delta$.}: $\Xi=\Xi_0 \pm \Delta$.  The meaning of this is depicted in Fig.~\ref{fig:disc}: if we measure the
property $Q$ --- which is what we can actually do, quoting it as a central value $Q_0$ and an errorbar $\Delta$ ---
and then translate the result in the $\Xi$ formulation, the interval of size $\Delta$ around $Q_0$ does not
correspond to a connected interval in $\Xi$ (or, if we insist to keep it connected, we loose precision in the
translation).  In the simple example described in \citep{Scorzato} we can even follow analytically how a connected
interval becomes disconnected.  But even without entering the details of a specific example, it suffices to observe
that nothing prevents the phenomenon described in Fig.~\ref{fig:disc}, and actually it is the minimum price we
should expect to pay when expressing the theory in the artificial formulation $\Xi$, that disregards the meaning of
any formula of the theory for the sake of conciseness\footnote{Note that we {\em can} associate a numerical value
  to $\Xi_0$, because a translation from $Q$ to $\Xi$ does exists. So, we cannot claim that $\Xi$ is not measurable
  in this naive sense.  The lack of measurability manifests itself only when considering the errorbars.}.

The impossibility to express the measurements in the form $\Xi=\Xi_0 \pm \Delta$ might seem a curiosity, but it has
actually profound consequences for philosophy of science.  In fact, this is a limitation of the $\Xi$ formulation
that we can {\em state precisely}, and for this reason we can exploit it for a precise formulation of scientific
theories, and for the characterization of at least one relevant cognitive value.

The idea is the following (see Sec.~\ref{sec:EEST}).  The postulates of any empirical scientific theory $T$ must
refer to at least a set $B$ of properties whose measurements are possible and can be reported as $b=b_0 \pm
\Delta$, for all $b\in B$.  Furthermore, the properties $B$ that appear in the postulates must be enough to enable
--- by possibly employing the laws of the theory --- the operative definition of all the other measurable
properties of $T$.  Now, the combination of these two requirements precludes a $\Xi$-like formulation of $T$.  As a
result, the formulation of $T$ cannot, in general be shorter than a minimal length, that is --- except for the
constraint of measurability --- {\em language independent} (see Sec.~\ref{sec:EEST}), and, hence, it is a well
defined property of $T$ (that we call {\em conciseness} of $T$).  By analogy with Kolmogorov complexity, such
minimal length is presumably not computable exactly, but can be estimated with finite precision.

Having defined a non-trivial notion of conciseness, it is natural to ask whether we can describe real scientific
progress as a Pareto improvement that takes into account only empirical accuracy and conciseness\footnote{We say
  that $A$ is {\em Pareto better} than $B$ for qualities $q_i$ ($i=1,n$), if $A$ is not worst than $B$ for any of
  the qualities $q_i$ and it is strictly better than $B$ for at least one of the qualities $q_i$.}.  I introduce
here a very simple model of progress along these lines and I contrast it with many possible objections and
challenging real cases of progress.

The plan of the paper is as follows.  In Sec.~\ref{sec:EEST}, I state precisely a property of direct measurements
that prevents $\Xi$-like reformulation, and I use it to introduce a minimal framework to represent scientific
theories and their admissible reformulations.  In Sec.~\ref{sec:MC}, I define the cognitive value of conciseness
and analyze some of its properties.  In Sec.~\ref{sec:SP}, I define a simple model of progress which is based only
on empirical accuracy and conciseness.  In Sec.~\ref{sec:ex} and \ref{sec:CP}, I test this model against
paradigmatic historic cases of real progress and against other classic philosophical views of progress.  Finally,
Sec.~\ref{sec:conclusions} contains my conclusions\footnote{Sec.~\ref{sec:EEST} and Sec.~\ref{sec:MC} summarize
  (and reformulate) the main content of \citep{Scorzato}, to make this paper self-contained.  Sec.~\ref{sec:SP} and
  Sec.~\ref{sec:CP} contain new material.}.

\vskip 0.3 cm

Before proceeding with the plan that I have just outlined, it is worth examining today's most popular alternative
approaches, in order to stress their limitations and unacceptable consequences.

One possibility, influentially advocated by \citet{Kuhn1} (see also \citealp{Laudan1981-LAUAPA}), is to rely on the
interpretation of these values by the {\em scientific community}.  Reliance on the scientific community would be
reasonable, if we could assume that the scientists rely, in turn, on well defined and durable values\footnote{I
  believe that this is indeed what good scientists mostly do, although often without recognizing such values
  wittingly.}.  But, since the possibility to characterize such values with any precision is, today, almost
universally denied, we can only hope that the scientists act rationally, even if nobody can tell why.  Should we
trust their future judgments on this issue, simply because they have solved some very different old problem or
because they were hired by someone who solved even older problems?  This can hardly be defended, since one of the
pillars of the idea of science is that even those scientists who produced the finest contributions can still do
mistakes, and they can be proven wrong even by the last student, if the latter provides better evidence.  In
science, it is always professed, it is not the authority --- and even less the opinion of the majority --- that
decides what is a good theory, but the evidence.  Emphasizing the role of the scientific community is simply a way
to elude the issue of investigating, seriously, the cognitive values on which the community relies, explicitly or
not.

An alternative view holds that the cognitive values are very clear to the experts, but they are {\em not general}:
they depend on the specific context or discipline and cannot be given a global definition (see, e.g.,
\citealp{Sober_PoS}).  But, if so, what does prevent any pseudo-scientists to claim that they are relying on
cognitive values that are actually well suited for the very specific context they are interested
in\footnote{Indeed, any present philosophical attempt to rebut even the most glaring cases of pseudo-science is
  fatally undermined by this weakness.  For example, some of the most insightful criticisms \citep{SoberID} to the
  Intelligent Design (ID) movement either contest the empirical accuracy of ID statements (which is certainly
  valuable, but powerless against cautious ID statements), or employ philosophical criteria, like falsifiability
  and its variations, that are contested by most philosophers.  In other cases, some criteria of demarcation
  between good and bad scientific practices are justified relying essentially on the {\em consensus} within the
  scientific community \citep{Leuschner-good-bad}.  Indeed, even in front of the most extreme cases of
  pseudo-science, we seem to have no reliable argument to tell why they do not fit in the history of science,
  without relying, for this conclusion, on a majority vote.}?  If we accept no general and well defined cognitive
value, we must also accept that {\em any} dispute over the scientific value of a method, or of a theory, may be
legitimately resolved with the foundation of a new discipline, suited to its very special context, with its own
scientific criteria, its own experts, its own journals, making up, overall, the greatest variety of incompatible
views, in which no one is better than any other.  I do not see much difference between denying the existence of
{\em general} and {\em well defined} cognitive values and claiming that {\em anything goes} \citep{Feyerabend}.

Many scientists perceive these problems as {\em only} philosophical, since they believe that valuable scientific
progress can still be recognized by its {\em applications}, especially technological ones.  Applications provide
indeed effective evidence of progress, as long as they flow copiously, even from the latest findings of basic
science.  But it is not always so, of course.  Nevertheless, modern research and education systems are based on the
belief that pursuing scientific knowledge itself eventually leads to much more useful results —-- in the long term
--- than just trying to improve what we think is useful now.  The reason why we need to identify cognitive, short
term, criteria for scientific progress is precisely because we cannot always rely on the immediate emergence of
useful applications, but we still need a way to assess the progress of our knowledge, that can plausibly lead to
new applications in the longer term.  If we want to use the amazing applications of Maxwell's and Einstein's
theories to justify the investments in modern research, we must be able to tell what is in common between what
Maxwell and Einstein did and what researchers do today.  Otherwise, what is expected to lead to useful outcomes in
the future?  Which (traditional or unorthodox) approach is more suitable to today's challenges?  In a rapidly
changing world, without a clear understanding of what has characterized science until now, it is not too difficult
for a careful charlatan to claim to be the authentic successor of Einstein.

But, even when the applications proliferate, reducing all measures of scientific progress to them is like reducing
science to a blind game of trial and error.  We know that science also has {\em internal} milestones: we do not
even dream about applications, unless we have already realized that we understood something significant.  What are
the considerations that, in those cases, tell us that we have improved our knowledge?  When scientists agree that a
new theory improves over another, they must use some criteria.  If we cannot even spell out clearly these criteria,
how can we claim that they are more valuable than anybody's taste?

Some decades ago, these issues were considered urgent, dramatic and painful by both the scientists and the
philosophers.  They stimulated heated debates, that culminated in the so called {\em science wars}
\citep{OneCulture}.  However, instead of prompting a careful and dispassionate assessment of the general cognitive
values actually used in science, that war was, in my opinion, little constructive, and the communication between
scientists and philosophers declined\footnote{An impressive evidence of this decline is offered by a recent review
  of a large sample of recent introductory scientific textbooks \citep{Blachowicz}, showing the predominance of
  very naive and long refuted views about the {\em scientific method} and the lack of nearly any influence from the
  last half century of philosophical analysis.}.  Today, we are left with a widespread belief that these issues are
unsolvable, but {\em only of philosophical interest}.  In this Introduction, I hope I have convinced the reader
that these issues cannot be dismissed as {\em only philosophical}: as scientists, we need to be serious about all
the cognitive values we employ, as much as we are about any key scientific concept.  Disregarding the role of
cognitive values is a serious threat to scientific progress.  In the rest of the paper, I argue that being serious
about non-empirical cognitive values is not impossible, especially if we regard this philosophical problem as a
scientific problem that requires the same scientific approach that we employ in any other scientific discipline.

\section{Empirical scientific theories and their reformulations}
\label{sec:EEST}

A prerequisite to discuss any cognitive value of scientific theories is to say what we mean by {\em scientific
  theories}.  This includes saying what may or may not count as a {\em valid reformulation} of a scientific theory,
since we have seen that $\Xi$-like reformulations undermine any attempt to express precisely any interesting
cognitive values.  The problem of a formal representation of scientific theories has been a matter of thorough
debate over the last century.  The new element that I introduce here is the realization that any {\em direct
  measurement} must be reported in the form $Q=Q_0\pm\Delta$, as discussed in the Introduction.  I argue that
taking into account this requirement in the formulation of scientific theories has deep implications.  In fact,
although it cannot {\em fix} the interpretation of the theory (nothing can do that), it does {\em exclude} some
interpretations that we have reasons to deem implausible.  This will be enough for the goals of this paper.  Let us
first state the requirement about direct measurements as follows\footnote{Postulate~\ref{def:ECDM} does not attempt
  to fully characterize the intuitive idea of {\em direct measurements}. In fact, Postulate~\ref{def:ECDM} might be
  fulfilled also by properties $Q$ that we would not regard as directly measurable.  But, for our purposes,
  identifying this property of direct measurements is sufficient. For this reason, this is introduced as a
  postulate and not as a definition of {\em direct measurements}.}:

\begin{post}
\emph{(Errorbar-connectedness of direct measurements).}
\label{def:ECDM}
The result of a valid direct measurement of a property $Q$ with central value $Q_0$ and inverse precision $\Delta$
is always expressed as a \underline{connected} interval\footnote{The detailed interpretation of this expression
  depends on the probability distribution $P(Q)$ that the theory assigns to each single measurement of $Q$
  (e.g. Gaussian, discrete...).  In this sense, Postulate~\ref{def:ECDM} rules out reports claiming that $P(Q_2) <
  P(Q_1)$ and $P(Q_2) < P(Q_3)$, if $Q_1<Q_2<Q_3$.  Note also that Postulate~\ref{def:ECDM} does not apply only to
  magnitudes assuming {\em real} values: $Q$ may represent any property that can be associated to a value in the
  course of an observation.  For example, in the context of a botanic theory, a typical observation may involve a
  decision whether an object is a tree or not.  In this case, the property {\em ``being a tree''} assumes values 1
  (={\tt true}) or 0 (={\tt false}).  I.e., it can be measured as much as the property {\em ``height of a tree''}.
  In all cases, the errorbar $\Delta$ remains meaningful and important, because the botanic theory, to which the
  concept of {\em tree} belongs, may need to account for the probability of failing to recognize a tree.  Hence,
  the theory must assign the proper meaning to $\Delta$, by associating to it a suitable probability of correct
  recognition.} as follows: $Q=Q_0\pm\Delta$.
\end{post}

Postulate~\ref{def:ECDM} is certainly a necessary condition for any property $Q$ that we might consider directly
measurable.  In fact, the result of any single direct measurement of any property $Q$ must be expressed as a {\em
  central value} and an {\em error-bar}, to have any plausibility.  On the other hand, Postulate~\ref{def:ECDM} is
also sufficient for the goals of this paper, because it makes the $\Xi$ trick impossible.  In fact, although an
imaginary supporter of the $\Xi$ formulation can always pretend to be able to interpret $\Xi$ by secretly measuring
$Q$ and then translating it into $\Xi$, he will not be generally able to quote the result as $\Xi = \Xi_0 \pm
\Delta$.

In general, the phenomenon of Fig.~\ref{fig:disc} is to be expected because the variable $\Xi$ is meant to encode
all the possible empirical consequences of the theory (being $\Xi=0$ its only law).  But, any sufficiently complex
scientific theory normally entails both consequences that can be measured very precisely, and consequences that
cannot be measured with any precision.  Interpreting $\Xi$ may be plausible only for toy-models where all the
possible consequences of the principles are immediately evident from the principles themselves, and one can either
measure everything with the same fixed precision or nothing at all.

Having justified Postulate~\ref{def:ECDM}, I now want to use it to characterize scientific theories to the extent
needed here.  The idea is the following.  Any empirical scientific theory is a mathematical theory, that must also,
somehow, make reference to some measurable properties.  The precise way in which such reference should be
formulated has always been controversial.  The minimalist approach adopted here consists in saying that (i) at
least some properties of the theory $T$ must be directly measurable, at least in the weak sense of
Postulate~\ref{def:ECDM}, and (ii) the measurements of any other measurable property of $T$ must be expressible in
terms of those that are directly measurable\footnote{Note that a theory typically contains also {\em
    non-measurable} properties, for which we put no constraints here.  Their role is important to improve the
  conciseness of the formulation. See Sec.~\ref{sec:MC}.}.  This idea is made precise by Def.~\ref{def:ST} and
Def.~\ref{def:MC}:

\begin{defn}
\emph{(Scientific theories).}
\label{def:ST}
A scientific theory is a quadruple $T=\{P,R,B,L\}$, where 
\begin{itemize}[noitemsep,nolistsep]
\item $P$ is a set of principles\footnote{The principles contain {\em all} the assumptions needed to derive the
  results of the theory, from the logical rules of deduction to the modeling of the experimental devices and of the
  process of human perception, so that no further {\em background science} is needed.  Note that also the {\em
    domain of applicability} of the theory can and must be defined by specifying suitable restrictions on the
  principles themselves.},
\item $R$ is a set of results deduced from $P$ (according to the logic rules included in $P$), 
\item $B$ is a set of properties that appear in $P$ and are \underline{directly measurable} in the sense of
  Postulate~\ref{def:ECDM} (we call them Basic Measurable Properties, or BMPs\footnote{For those familiar with
    \citep{Scorzato}, which is {\em not} a prerequisite for this paper, note that I adopt here a simplified
    notation with respect to \citep{Scorzato}, where the BMPs (= BECs) were not necessarily included in the
    principles of $T$, but were included in the string $\sigma$ relevant for the evaluation of the complexity of
    $T$.  It seems more convenient to include the BMPs directly in the principles, as done here.}, of $T$),
\item $L$ is the language in which all the previous elements are formulated.
\end{itemize}
\end{defn}

Note that Postulate~\ref{def:ECDM} and Def.~\ref{def:ST} cannot fix the interpretations of the BMPs.  Nothing can
do that: theories face the tribunal of experience as a whole \citep{Quine2Dogs}, and the assumptions of sufficient
unambiguity of their BMPs are necessarily part of the theoretical assumptions.  The aim of Postulate~\ref{def:ECDM}
is not to fix the interpretation of any theoretical expression, its aim is rather to exclude a class of
interpretations that we deem certainly implausible.

Besides the BMPs, a theory can typically define many other (possibly unlimited) measurable properties.  These can
be characterized as follows:

\begin{defn}
\emph{(Measurable properties).}
\label{def:MC}
The measurable properties (MPs) of a theory $T$ are all those properties that can be determined through
observations of the BMPs $B$ of $T$, by possibly employing some results $R$ of $T$.  Their precision is also
determined by $T$.
\end{defn}

Hence, the BMPs must be sufficient to enable the measurements of all the MPs that the theory needs to
describe\footnote{The demand that a sufficient amount of BMPs be included in the principles should not be confused
  with the much stronger demand that a theory should be formulated in terms of MPs {\em only}, as wished by
  Heisenberg \citep{Born:1961:BSD}, but never realized.  Non-MPs, such as the {\em quark}'s wave functions, are
  often essential to formulate a theory concisely, but the theory {\em also} needs to refer to a sufficient amount
  of MPs.}.  In other words, the BMPs provide --- together with the principles to which they belong --- the
basis\footnote{It is not a {\em universal} basis as in \citep{CarnapAufbau}.  All MPs (basic or not) are completely
  theory dependent.} on which the whole interpretation of the theory is grounded.  Thanks to the identification of
the BMPs, the principles truly encode {\em all the assumptions} of the theory, in a sense that goes beyond the
logical structure of the theory.  This observation deserves to be emphasized:

\begin{rmrk}
The principles $P$ of a theory $T$ encode all the information needed to fully characterize $T$, in the following
sense: the $P$ are sufficient, in principle, to enable anyone to check whether any given derivation of a result
$r\in R$ is actually valid in $T$.  Moreover, the principles $P$ are sufficient to enable anyone who can interpret
the BMPs $B$ to check the agreement between any result $r\in R$ and any experiment.
\end{rmrk}

We can finally address the question that motivated us at the beginning of this Sec.~\ref{sec:EEST}: to what extent
can we claim that a theory $T'$ is only a reformulation of another theory $T$? According to Def.~\ref{def:ST} any
translation of $T$ in a different language counts as a different theory.  But we obviously want to identify
different formulations, as long as their differences are not essential.  This is the case when two theories are
equivalent both from the logical and from the empirical point of view, i.e., when all their statements concerning
any MPs agree.  More precisely:

\begin{defn}
\emph{(Equivalent formulations for $T$).}
\label{def:EL}
We say that $T$ and $T'$ are equivalent formulations iff:
\begin{itemize}[noitemsep,nolistsep]
\item[(i)] there is a translation ${\cal I}$ between $T$ and $T'$ that preserves the logical structure and the
  theorems\footnote{Specifically, I refer to what is called bi-interpretability \citep{Visser1991}.  However, other
    notions of logical equivalence would not change the present discussion, since they cannot, alone, rule out a
    translation into a $\Xi$-like formulation.} ({\em logical equivalence});
\item[(ii)] and for each MP $c$ of $T$ (resp.~$c'$ of $T'$), ${\cal I}(c)$ (resp.~${\cal I}^{-1}(c')$) is also
  measurable with the same precision and the same interpretation\footnote{I.e., an experiment that measures $c$
    within $T$ also measures $c'$ within $T'$.} ({\em empirical equivalence}).
\end{itemize}
${\cal L}_T$ denotes the set of all pairs $(L,B)$ of available languages and BMPs in which we can reformulate $T$
and obtain a new theory $T'$ that is equivalent to $T$.  In the following, the symbol $T$ refers to a scientific
theory up to equivalent formulations, while $T^{(L[,B])}$ refers to its formulation in the language $L$ [and basis
  $B$]\footnote{Note that we do not require that $B'={\cal I}(B)$: two equivalent theories may choose different
  BMPs, because what is basic for one theory may not be basic for the other.  Only the equivalence of the MPs is
  required.}.
\end{defn}

In particular, Def.~\ref{def:EL} implies that the $\Xi$ formulation is not equivalent to the Standard Model: the
translation that makes them logically equivalent cannot realize also an empirical equivalence, because the $\Xi$ is
not an acceptable MP, for the Standard Model.  This is how the $\Xi$ trick is evaded.

\subsection{Example: classical electromagnetism}
\label{sec:ex1}

For example, consider classical electromagnetism, that can be defined, primarily, by the microscopic Maxwell's
equations (in convenient units and standard notation \citep{cottingham2007introduction}):

\begin{equation}
\label{eq:max}
 \nabla\cdot{\bf E} = \rho 
\qquad
 \nabla\cdot{\bf B} = 0  
\qquad
 \nabla\times{\bf E} = - \frac{\partial{\bf B}}{\partial t} 
\qquad
 \nabla\times{\bf B} = {\bf j} + \frac{\partial{\bf E}}{\partial t}.
\end{equation}

But the theory is not completely defined without specifying which ones of the properties involved can be measured:
{\em space} and {\em time} coordinates of the charges, the magnitudes of ${\bf E}$, ${\bf B}$, ${\bf j}$, ${\bf
  \rho}$.  For example, we could use an {\em electric field meter} to measure ${\bf E}$ and a {\em magnetometer} to
measure ${\bf B}$, within some ranges.  We do not need to enter any detail of the functioning of these tools and
the physical laws behind them, to define our theory precisely.  We simply need to assume that such tools are
available, and they relate the values of ${\bf E}$ (resp.~${\bf B}$), within some ranges, to some numbers in their
digital displays.  We can assume as BMPs of our theory either the reading of the digital displays or directly ${\bf
  E}$ and ${\bf B}$.  The same can be repeated for the other quantity entering Eq.~(\ref{eq:max}).  Note that we
have to {\em assume} that the BMPs are interpreted unambiguously by any observer.  We cannot prove that this must
happen, but if it does not, it counts as lack of empirical accuracy of any theory containing these theoretical
assumptions.

\section{A measure of complexity of the assumptions}
\label{sec:MC}

After having introduced a representation for scientific theories and their admissible reformulations, I turn to the
original goal of identifying at least one non-empirical cognitive value that can justify theory selection.  As
noted in the Introduction, non-empirical cognitive values are necessary to penalize those aspects of scientific
theories that cannot be penalized by empirical accuracy alone: these are ad-hoc hypothesis, the addition of many
extra-terms in models, or long list of exceptions.  Can we penalize these without penalizing valuable theories?

Having ruled out the $\Xi$ trick, we are in the position to attempt something new.  In fact we can define the
complexity of a theory $T$ as the {\em minimum} over the {\em truly equivalent} formulations of $T$:

\begin{defn}
\emph{(Complexity of the assumptions; conciseness).}
\label{def:CT}
Let $P^{(L,B)}$ be the principles of $T$, when expressed in language $L$ and with BMPs $B$.  Let the {\em
  complexity of the assumptions} of $T$ be:
\begin{equation}
\label{eq:CT}
{\cal C}(T) = \min_{(L,B) \in {\cal L}_T}  {\rm length}[P^{(L,B)}]
\end{equation}
Let the {\em conciseness} of $T$ be the inverse of ${\cal C}(T)$: ${\rm Conc}(T)=1/{\cal C}(T)$.
\end{defn}

Note that, if ${\cal L}_T$ had included all logically equivalent formulations, regardless of empirical equivalence,
the minimum would have been trivial, because $\min_{(L,B)} {\rm length}[P^{(L,B)}] = {\rm length}[\Xi] = 1$.  But
now, improving the conciseness of a formulation cannot be done without ensuring a connection between the principles
of the theory and truly measurable properties. Hence, ${\cal C}(T)$ is not trivial in general.

On the other hand, Def.~\ref{def:CT} effectively assigns to $T$ that notion of complexity in which $T$ fares best,
under no other constraint than measurability, as stated in Postulate~\ref{def:ECDM}.  {\em I conjecture that
  Def.~\ref{def:CT} represents --- within the limited precision associated to it --- a cognitive value able to
  jusitfy scientific theory selection, when combined with empirical accuracy}.  The rest of this paper is devoted
to support this conjecture.

\subsection{Analysis of ${\cal C}(T)$}

The proper justification of the above conjecture can only come from the comparison with real cases of scientific
theory selection\footnote{In this sense I see philosophy of science as an empirical science itself, whose goal is
  understanding the rules behind that historical phenomenon that we call science.}.  But, before doing that in
Sec.~\ref{sec:SP} and Sec.~\ref{sec:CP}, I comment on some general aspects of ${\cal C}(T)$:

1.  The first comment concerns the accessible precision in computing ${\cal C}(T)$.  Although ${\cal C}(T)$ is well
defined and non-trivial, it is certainly very hard to compute in practice\footnote{And perhaps even impossible to
  compute in principle, because of its likely relation with Kolmogorov $K$ \citep{Chaitin}.  As opposed to ${\cal
    C}(T)$, the Kolmogorov function $K$ is defined for a fixed language and a fixed computing model
  \citep{LiVitanyi1997}.  This dependence on the language makes $K$ unsuitable for direct application in philosophy
  of science, as discussed in \citep{Kelly-razor}.  The definition of ${\cal C}(T)$ removes the language/model
  dependence by taking the minimum over all formulations that fulfill measurability constraints, as justified in
  the text.  This difference makes it hard to extend the proof of non-computability to ${\cal C}(T)$.  But what
  matters here is that ${\cal C}(T)$ is certainly hard to compute.}.  Even though we are always only interested in
comparing the complexity of two alternative theories, that can be expressed by $\delta {\cal C}(T,T') := {\cal
  C}(T) - {\cal C}(T')$, also $\delta {\cal C}$ is often very hard to compute and can be estimated only
approximatively.  Often, we are not even able to tell whether $\delta {\cal C}$ is positive or negative.  In fact,
modern scientific theories typically combine many assumptions from very different scientific fields.  Even when all
the assumptions that distinguish $T$ from $T'$ are clearly identified, finding the formulations that minimize
respectively ${\cal C}(T)$ and ${\cal C}(T')$ may require rewriting a substantial part of the body of science.  For
this reason, we must often rely on an estimate based on the traditional formulation.  Moreover, in some cases, the
full list of the assumptions of a theory is not entirely specified.  This may happen, for example, when a theory is
considered in a preliminary form by its very proponents (a status that may last a long time); but it may also
happen when old theories contain implicit assumptions whose necessity was overlooked, until an alternative theory
reveals them (see Sec.~\ref{sec:darwin}). All this adds further uncertainty on the estimate of $\delta {\cal C}$.

But the limited precision of $\delta {\cal C}$ is exactly the feature that we expect from a sensible notion of
complexity in science.  Because, in scientific practice, we do not rely on complexity to discriminate between
theories with a roughly similar amount of assumptions, since we know that some overlooked formulation might easily
reverse our assessment.  In those cases, we need to suspend the judgment on simplicity (i.e. accept $\delta {\cal
  C}\simeq 0$, within errors) and rather look for potential different predictions.

On the other hand, there are also many important cases where it is totally unambiguous that $T$ is simpler than
$T'$.  This is especially important when $T'$ achieves good accuracy only because it puts little effort toward any
synthesis.  This is the case, for example, when $T'$ adds new parameters or ad-hoc assumptions; or when $T'$ is
built by patching together different models, each capable of describing only a small subset of the data; or, in the
extreme case, when $T'$ is just a collection of experimental reports.  In these cases, the scientists often do not
even consider $T'$ a {\em theory}, but this can be justified only because they use --- {\em implicitly} but {\em
  essentially} --- a notion equivalent to $\delta {\cal C}$ to rule out $T'$.

This picture is fully consistent with the intuitive idea that the notion of complexity is ambiguous, but only to
some extent, because there are many cases in which there is absolutely no doubt that $T$ is simpler than $T'$, in
any conceivable and usable language.  This {\em limited precision} without {\em arbitrariness} cannot be justified
by a generic appeal to different opinions. But it does find justification in computational limitations (under
constraints of measurability) of a well defined notion of complexity.

2.  The second comment concerns possible alternatives to Def.~\ref{def:CT}. In particular, one may argue that the
usage of the {\tt length[]} in Eq.~(\ref{eq:CT}) is just one arbitrary choice.  However, since the minimum is taken
over all possible formulations, I argue that Def.~\ref{def:CT} effectively takes into account any {\em plausible}
notion of complexity.  For example, instead of the function {\tt length[]}, one might assign more weight to some
symbols, or combinations of symbols.  But this would be equivalent to a formulation in which those (combinations
of) symbols are repeated, and we still use the {\tt length[]}.  Hence, this possibility is already included in
Def.~\ref{def:CT}, but it is not minimal.  Alternatively, one might wish to count only some kind of symbols
(i.e.~give zero weight to others)\footnote{Any notion of complexity based on counting entities or quantifiers would
  fit in this case.}.  But if neglecting symbols that we cannot reduce reverses a decision about which theory is
better, it is hard to claim that these symbols should not count!  One can of course consider any other function of
$P^{(L,B)}$, but, when it is too different from any traditional notion of complexity --- that inevitably boils down
to counting something --- it becomes very difficult to justify it as a plausible notion of complexity.
%% ch

These arguments do not intend to justify Def.~\ref{def:CT} {\em a priori}.  Def.~\ref{def:CT} can only be justified
by showing that it reproduces the preferences that we expect in paradigmatic real cases.  To challenge my
conjecture, one should find at least one case where our best estimate of $\delta {\cal C}(T,T')$ for empirically
equivalent $T$ and $T'$ gives an intuitively unacceptable result.
%% ch

Combining observations 1. and 2. leads to a new, very important, observation: two different definitions of
complexity, that are nevertheless consistent within errors, entail identical consequences, and it is immaterial to
discuss which one we may prefer.

3.  The third comment concerns the restriction in Def.~\ref{def:CT} to {\em available} languages.  There is no
doubt that ${\cal C}(T)$ strongly depends on the set of languages that we are able to conceive, but finding simpler
formulations is hard.  Hence, this constraint does not represent a bias of our characterization of complexity, but
a limit of our {\em knowledge}, which is exactly what ${\cal C}(T)$ needs to account for.  If we feel that the
assumptions of $T$ are simpler than those of $T'$, but we are unable to formulate $T$ such that ${\cal C}(T) \leq
{\cal C}(T')$, it must be really unclear to us in which sense we expect the assumptions of $T$ to be {\em simpler}.
If we finally succeed, the new formulation should be regarded as a progress of our knowledge.

4.  The final comment concerns the suspect of a possible bias due to the particular condition on direct
measurements set by Postulate~\ref{def:ECDM}.  This might cause difficulties only if a scientific theory appears
simpler when formulated in terms of BMPs that do fulfill Postulate~\ref{def:ECDM}, but that are not {\em really}
measurable.  Should this be the case for a (relevant) scientific theory, it would force us to refine our condition
on direct measurements in Postulate~\ref{def:ECDM}.  But, until this happens, no bias can come from it.

\section{A model of scientific progress: describing more with less}
\label{sec:SP}

Having formulated a notion of minimal complexity of the assumptions ${\cal C}$, in Def.~\ref{def:CT}, we can
combine it with the notion of empirical accuracy\footnote{In this paper we do not need more than the intuitive
  notion of empirical accuracy.  However, for completeness, a definition is proposed in App.~\ref{sec:app}.  We
  always assume that empirical accuracy refers to an unspecified fixed set of MPs.} to give a tentative meaning to
the notion of {\em better theories}.  This leads to a simple model of {\em scientific progress}, which is based
only on these two cognitive values.  Here, I first define the model, while the rest of the paper is devoted to test
how the model fares with respect to paradigmatic real cases of scientific progress and with respect to other views
of progress\footnote{The tests presented here are only a first step, but fully in the spirit of
  \citep{Laudan1986-LAUSCP}.}.

Since the role of empirical accuracy in theory selection is undisputed, comparing real cases of progress to this
model is the proper way to test the conjecture that the complexity of the assumptions ${\cal C}$ represents well
the non-empirical cognitive values that actually matter in science.

\begin{defn}
\emph{(Better theories; state of the art; outdated theories; scientific progress).}
\label{def:SP}
Let a theory $T$ be {\em better} than $T'$ if $T$ is more empirically accurate or has lower complexity of the
assumptions than $T'$, without being inferior in any of these aspects.  If there is a theory $T$ better than $T'$,
we say that $T'$ belongs to the {\em outdated theories}.  Otherwise, we say that $T'$ belongs to the {\em state of
  the art}.  Finally, we say that there is {\em scientific progress} when a state of the art theory $T$ becomes
outdated\footnote{Note that this can only happen because either a new theory $T'$ appears, that is better than $T$,
  or because a new experiment causes an existing theory $T''$ to become better than $T$.}.  We call this model of
scientific progress $SP_0$.
\end{defn}

\begin{figure}
\centering
\includegraphics[width=120mm]{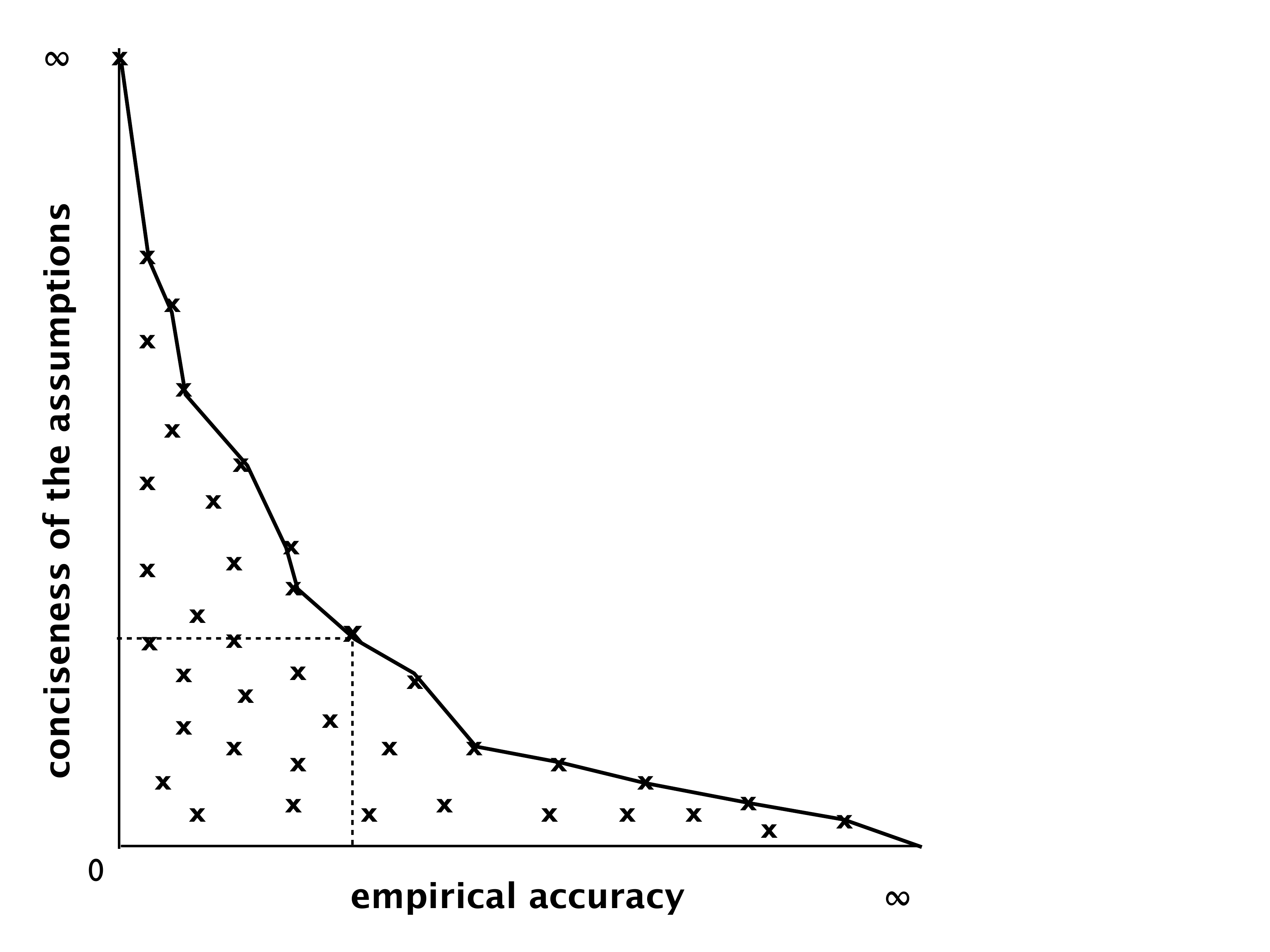}
\caption{We plot scientific theories (crosses) according to the simplicity of their assumptions and their empirical
  accuracy (the latter is actually multi-dimensional, although only one dimension is shown here).  The thick line
  joins the {\em state of the art} theories, while the {\em outdated} theories lie below that line.  The dashed
  square contains the theories that are worse than the theory in the top right corner of the square.}
\label{fig:pp}
\end{figure}

The notion of scientific progress defined above may be visualized with the help of Fig.~\ref{fig:pp}.  Note that
the state of the art may include also theories that are extremely simple but empirically empty (e.g. those at the
top left of Fig.~\ref{fig:pp}) and theories that are extremely lengthy but agree very well with the experiments
(e.g. the collection of experimental reports, at the bottom right of Fig.~\ref{fig:pp}).  We have no unambiguous
way to exclude them from the state of the art (and probably we should not).  Nevertheless, Def.~\ref{def:SP} is
able to justify the growth of scientific knowledge \citep{PopperCR}, in the sense that very popular scientific
theories are regularly overthrown and superseded by better ones.  {\em Moving the edge of the state of the art is
  what constitutes scientific progress}, and this is what valuable science does, quite often.  But, it does not
achieve it trivially: for example, collecting more and more experimental reports with unsurprising results, does
not make any old theory outdated, and it does not produce, by itself, progress.  For the same reason, we do not
make progress by calculating quantities that we cannot measure or by fitting the data thanks to more parameters.

Note that to each cross in Fig.~\ref{fig:pp} we should attach an errorbar in both axes.  This makes every statement
about empirical accuracy, simplicity, better theories, etc., a provisional one.  For example, new experiments, or a
better estimate of simplicity, may bring back to the state of the art an already outdated theory.  This is always
possible, in principle.  The errorbars tell us how unlikely we estimate such event to be.  This should not
surprise: like any good scientific concept, also the philosophical concept of {\em scientific progress} can be {\em
  precisely defined} even though the assessment of its occurrence is necessarily {\em approximate} and {\em
  revisable}.

The state of the art represents, as a whole, our current scientific image of the world \citep{SellarsSI}.  The
theories that belong to it cannot be assembled in a single, logically coherent, best theory of all.  But they
represent, altogether, the toolkit from which we can chose the tool(s) that are best suited to the given empirical
questions and to the given requirements of accuracy\footnote{There can be more than one theory with the same
  complexity and the same empirical accuracy (but with different untested predictions).  Moreover, as said, the
  state of the art is known only approximatively.  The standard way to formulate predictions from this collection
  of approximatively state of the art theories is a Bayesian approach \citep{Bayesian}, combined with some choice
  of trade-off between simplicity and accuracy.  Note that no trade-off is necessary to justify progress: it is
  only needed to produce a single prediction out of the chosen subset of the state of the art.}.  Some theories
based on Newton mechanics still belong to the state of the art for those issues where quantum mechanical effects
are undetectable or where the relevant results cannot yet be deduced from a more fundamental set-up.  Moreover,
when we are overwhelmed by surprising experimental results, in which we cannot find any regular pattern, even the
collection of all experimental reports may be the best theory we have (see Sec.~\ref{sec:ex-qm}).

This picture has also important implications concerning the proper way of {\em communicating scientific results}.
In fact, statements like {\em ``Phenomenon $X$ is a scientific fact''} (independently of any assumption) are never
fully justifiable.  Too often they needed to be retracted, which contributed to discredit the scientific method.
They represent bad communication at best.  But the alternative is not anarchy, because statements like {\em ``All
  the state of the art theories that are able to describe the set $S$ of phenomena they also imply phenomenon
  $X$''} retain, instead, full philosophical and scientific justification.  Although the latter might sound weaker,
all their important practical implications are exactly the same.  Indeed, when a good scientist confronts a claim
that denies phenomenon $X$, she will not appeal to her authority, she will rather asks: ``What is your alternative
theory? Is it able to account for basic phenomena? Does it introduce unnecessary ad-hoc hypothesis?''.  This
clearly reveals an implicit thinking in terms of the state of the art.

Although I have stressed the important role of the state of the art, outdated theories are not thrown away, since
hardly anything is thrown away in science.  They might still contain ideas that will eventually prove fruitful in
the future.  But we would never use them in any application, and we would never attribute any value to their
predictions.

I address now some possible challenges to the model $SP_0$ of Def.~\ref{def:SP}.  First, one might suspect that
$SP_0$ fails to recognize the progress that is achieved when a great improvement of empirical accuracy is obtained
a the cost of a slight increase of complexity.  The example in Sec.~\ref{sec:ex-qm} illustrates why this is
practically never the case.  Second, one might think that there can be progress even when no theory becomes
outdated.  This might be possible, but the example in Sec.~\ref{sec:limit} suggests that this is rarely the case.
Third, does $SP_0$ apply only to highly mathematical theories?  The examples in Sec.~\ref{sec:darwin} and
\ref{sec:ssh} provide evidence to the contrary.  Finally, one might think that a theory $T$, more complex than
$T'$, and empirically equivalent to it, might have other cognitive advantages.  This possibility is discussed in
Sec.~\ref{sec:ocv}.

\subsection{Examples and challenges}
\label{sec:ex}

\subsubsection{Electromagnetism again}
\label{sec:ex-sp}

As a first example, compare Eq.~(\ref{eq:max}) with their relativistic covariant expression\footnote{See, e.g.,
  \citealp{cottingham2007introduction}, Sec.~4.1, for the derivation of Eq.~(\ref{eq:max}) from
  Eq.~(\ref{eq:cov-max}).}:
\begin{equation}
\label{eq:cov-max}
\Box A  = j
\qquad
\partial \cdot A = 0
\end{equation}
The advantage of the covariant formulation is not so much its conciseness, in fact, $A=(A^0,{\bf A})$ is not
measurable, and we still need to define ${\bf E} = -\nabla A^0 - \frac{\partial {\bf A}}{\partial t}$, \;${\bf B} =
\nabla \times {\bf A}$, and $j=(\rho,{\bf j})$, to find suitable BMPs.  The real advantage of
Eq.~(\ref{eq:cov-max}) is that, {\em without appreciable loss of simplicity}, it is now directly applicable to
different reference frames (being automatically Lorentz invariant), and hence it covers a much wider range of
experiments.

\subsubsection{Quantum mechanics}
\label{sec:ex-qm}

An example that might appear problematic for $SP_0$ is the introduction of quantum mechanics.  One may argue that
quantum mechanics achieves more empirical accuracy than classical mechanics only at the expenses of a higher
complexity of the assumptions.  This is true, nevertheless, quantum mechanics is certainly better than the
collection of experimental reports that described the new surprising quantum phenomena in the early 20th century,
and it is also better than those early phenomenological laws that were devised to describe only specific quantum
systems\footnote{Namely, Planck's theory of the black body, Einstein's theory of the photoelectric effect, Bohr's
  model of the atom, etc.}.  Some combination of the latter was in the state of the art, until it was overthrown by
the introduction of quantum mechanics in its final form, which, therefore, did produce progress.

The lesson of this example (and the previous one) is the following.  If we focus only on two subsequent {\em
  famous} theories, we often see a big increase in empirical accuracy, with a change in the complexity of the
assumptions that may be difficult to estimate (or might even worsen).  This limited perspective suggests that a
trade-off between simplicity and empirical accuracy has occurred.  But the measure ${\cal C}(T)$ is crucial to
recognize, unambiguously, the superiority of the new famous theory not necessarily vs the famous older one, but vs
the patchwork theories that represent early attempt to fit together the new experimental data through various
half-baked models.  It is not always the older famous theory which is made outdated, but those, less famous,
intermediate theories.  Hence, it is on the larger picture of Fig.~\ref{fig:pp} that ${\cal C}(T)$ plays its less
obvious but essential role to identify unambiguous progress, with no trade-off.

\subsubsection{General relativity and the bending of light}
\label{sec:limit}

Another possible objection is that progress might also occur without any state of the art theory becoming outdated.
I cannot exclude this in principle, but I am not aware of any important scientific result that did not make any
theory outdated.  

One case that is apparently very challenging in this respect is the early confirmations of Einstein's theory of
gravity (GR).  One might argue that, after GR successfully accounted for the precession of Mercury, it was already
better than Newton's theory (NG), to describe gravitational phenomena, and the impressive prediction of the bending
of light in 1919 did not bring further progress, according to $SP_0$, which sounds very counter-intuitive.  But,
this account is not complete.  Consider the theory NG' defined by: NG + electromagnetism (EM) + the assumption that
the aether is bound to the earth + the assumption that the precession of Mercury was due to some other unknown
effect.  In 1919, NG' was as empirically accurate as GR'=GR+EM.  Moreover, the complexity of NG' was not yet
clearly worse than that of GR'.  Indeed, the very short list of ad-hoc assumptions we have introduced to define NG'
was sufficient to save the very few anomalies that NG manifested in 1919.  Moreover, the mathematics behind GR' was
not so much used in physics, and {\em that} could be legitimately considered rather ad-hoc\footnote{Actually, it
  was not so for the very few people, like Einstein, who appreciated the great conciseness of the principles behind
  GR.  He probably estimated NG' clearly worst than GR' already before 1919.  Indeed, it is well known that
  Einstein was not so impressed by the experiment of 1919 \citep{Rosenthal-Schneider1980}.  For him, that
  experiment was, perhaps, almost a routine confirmation.  This confirms that the uncertainty in our estimate of
  ${\cal C}$, agrees with the uncertainty that we should retain in assessing whether there was actually significant
  progress.}.  The discovery of the bending of light made NG' unambiguously worse than GR' and outdated.

As soon as GR accumulated more and more evidences, further corrections of NG became more and more
cumbersome\footnote{This gradual process was seen by \citet{Kuhn1} as a social process of conversion.  However, it
  is completely understandable in terms of increasing complexity of the alternative theories, that only gradually
  exceeds the estimated errorbars.}.  Indeed, today, a possible experimental confirmation of gravitational waves,
would not by itself represent progress, according to Def.~\ref{def:SP}, unless such discovery also rules out some
alternative state of the art theory\footnote{Indeed, we always consider much more interesting those experiments
  that contradict well established theories, rather than those confirming them.}.

For these reasons, the present characterization of progress is perhaps incomplete.  But, some incompleteness is
probably the price to pay for any characterization that only includes genuine progress.  Moreover, it applies to a
very large amount of real historical cases, and hence, it can be regarded as a challenging, but realistic goal for
the scientific community.

\subsubsection{Darwin's theory}
\label{sec:darwin}

Some scientific theories are believed to have principles that cannot be used to {\em formally deduce} observational
statements, but they should be interpreted more {\em informally}, as sources of explanations.  For example,
\citet{Kitcher1993-AOS} remarks that the principles that he extracts from \citep{Darwin-origin}\footnote{They are
  the principles of {\em (1) variation, (2) struggle for existence, (3) variation in fitness, (4) inheritance}.  As
  stressed by Kitcher, these principles were not controversial: they could be considered part of well accepted
  theories.}  are not sufficient to deduce any important conclusion of Darwin's theory.  Instead, he introduces the
concept of {\em Darwinian histories} that provide bases for acts of explanations that are linked by analogy of
patterns, rather than deductively.

Without denying the interpretative value of Darwinian histories, another, more classic and deductive account of the
achievements of Darwin's theory is certainly possible.  For this, we need to recognize another crucial assumption
of Darwin's theory (although much less emphasized): the fact that all the species that he considers have a common
ancestor \citep{SoberDarwin}.  This assumption is obviously not sufficient to deduce the details of how any specie
may evolve, but it is sufficient to deduce some key features of the probability distribution of the species across
geographical space and geological time.  In particular, Darwin's model predicts that slightly different species are
more likely to be found close in space, and gradually different species are expected to be found in successive
fossil records.  Moreover, Darwin's theory predicts a high probability to find, across different species, common
features that are not justified by fitness.  Although Darwin does not use the language of probability
distributions, and he was not in a position to deduce any quantitative estimate, the core of {\em The Origin} is
devoted precisely to discuss such (qualitative) predictions, and compare them with the available evidence.  Darwin
emphasizes that, in order to reproduce the observed geographical and geological correlations, a non-evolutionary
theory needs to postulate a very complex structure of separate acts of creation. On the other hand, Darwin's
assumptions implied precisely the kind of correlations that Darwin gathered in his research.

Did Darwin's theory achieve progress? It certainly did with respect to Lamarck's theory, that entailed similar
consequence as Darwin's but at the cost of the additional assumption of two totally new forces: a {\em
  complexifying force} and an {\em adapting force}.  Such forces were not necessary for Darwin, who relied only on
well accepted basic mechanisms, since he realized that nothing {\em required more} than those.

It is more difficult to assess whether Darwin's theory was progressive with respect to a creationist one.  This is
because there is a great variety of possible creationist theories.  It is of course impossible to beat the
empirical accuracy of a theory that assumes the creation of each single specie as it is, in the moment it is first
found.  But, in the early 19th century, it was clear that even the monumental Linneaus' work was very far from
providing a complete list to a similar theory.  Alternatively, one could at least postulate the {\em rules}
according to which the centers of creation were distributed, without specifying the details of each center and each
specie.  This is essentially the kind of explanation that Darwin was challenging his opponents to provide (although
using a different language).  In this sense, Darwin's theory was progressive with respect to a family of possible
theories to which many people referred, but no one could formulate explicitly.

These brief notes certainly do not represent a complete analysis of Darwin's theory in the context of the model of
progress proposed here.  But they show that our analysis is certainly applicable well beyond highly mathematical
disciplines\footnote{In this context, the role of measurability lies behind the scene, but it is essential to force
  us to express Darwin's theory by means of traditional concepts from biology, geology, physics, chemistry, etc.
  In fact, Darwin's theory is meaningful only in combination with a sufficiently rich background of standard
  disciplines.}.

\subsubsection{History is a science}
\label{sec:ssh}

The model $SP_0$ is also not restricted to natural sciences.  For example, the task of an historian is to
reconstruct a series of events that cannot be checked directly, but whose likely consequences explain the available
documents.  In general, no document or collection of documents can {\em prove} anything, without additional
assumptions, just like no experiment can {\em prove} a scientific theory.  A careful use of the laws of probability
can only make some reconstructions more likely than others, under suitable assumptions.  The question is {\em which
  assumptions}.  Ideally, one would like to rely only on {\em well established} natural laws, which would leave all
the burden of discussing non-empirical cognitive values to the natural scientists.  But, as all historians know too
well, natural laws are far from sufficient to solve historic issues, even if we content ourselves with
probabilistic answers.  For example, many important historic questions amount to ask the {\em motivations} why,
say, king Henry did something.  Such questions are not less scientific than questions about king Henry's date of
death, and the historians' answers is often neither less scientific nor less convincing than many explanations by
physicists.  But the answer to these questions necessarily rely on different assumptions.  Much evidence about
previous acts and experiences by king Henry can make one explanation (much) more likely than another, but, at the
end, any answer must rely also on some implicit or explicit model of what are the possible and likely human
motivations and preferences under general circumstances.  Psychologists and sociologists can collect many data to
support one model over another, but they cannot consider all the possible circumstances and even the best model
must assume that some parameters are irrelevant, even if nobody has tested that claim.  The choice to neglect some
possible parameters is the crucial choice that is justified only in terms of simplicity, which, in turns, cannot be
justified in terms of more fundamental laws.  Furthermore, the kind of simplicity that matters here is in fact a
form of conciseness, because omitting the dependency of some parameters is clearly the most concise choice (in
ordinary language)\footnote{Of course, behind the decision to neglect some parameters there is always the hope that
  they are indeed irrelevant.  But, in absence of any evidence in favor or against, any choice different from the
  most concise, in some formulation, would be considered ad-hoc.}.  Also in this case, the role of measurability is
that of justifying the use of ordinary language.

\section{A consistent picture}
\label{sec:CP}

In this last section I compare the model $SP_0$ with other classic views of progress and goals of science
(Sec.~\ref{sec:ogos}).  Later, I discuss whether other cognitive values, besides empirical accuracy and the
complexity ${\cal C}$ of the assumptions, need to be part of an adequate description of scientific progress
(Sec.~\ref{sec:ocv}).

\subsection{Other views of progress and/or goals of science}
\label{sec:ogos}

First of all, we should note that $SP_0$ covers all cases of genuine {\bf reduction}.  In fact, all models of
reduction described by \citet{reductionism-IEP} imply that, after showing that the theory $T$ reduces to the theory
$B$, we can describe all the phenomena previously described by the union of $T$ and $B$ with strictly {\em less}
assumptions, namely those of $B$ alone.  More precisely, let $B^{\rm new}$ be the new theory that has the same
assumptions as $B$, but contains the new results, namely those that describe all the phenomena previously described
by $T$.  This means that $B^{\rm new}$ makes $T + B$ outdated, since $B^{\rm new}$ becomes at least as empirically
accurate as $T + B$, but it is simpler, since it has strictly less assumptions (only those of $B$).

Actually, we did not need a notion of complexity to characterize the progress due to reductions, since, in this
case, the assumptions of $B$ are a {\em subset} of those of $T + B$.  In fact, a satisfactory characterization of
progress in these cases already existed.  But the advantage of $SP_0$ is that, besides subsuming the classic cases
of reductions, it also accounts for episodes of genuine {\bf revolutionary theory change}, as described in
Sec.~\ref{sec:ex}.

On the other hand, if we want to compare $SP_0$ with other views that try to account for progress {\em beyond
  reductions}, we face the problem that all those views {\em need a notion of complexity to be defined}.  I argue
that, if we use the notion of complexity ${\cal C}$, to fix those definitions, we obtain well defined models of
progress that are consistent with their original motivations and are consistent with $SP_0$.  A detailed analysis
would require definitely more space than it is possible in this article, but here I can at least sketch the key
ideas.

Successful {\bf predictions} are often regarded as classic landmarks leading to the acceptance of a new theory.  In
order to be useful for theory choice, predictions should be able to discriminate between different theories. But,
this is not sufficient: predictions can be valuable only if they are based on {\em valuable} theories and they rule
out {\em valuable} competing theories (otherwise they simply amount to routine checks).  On the other hand,
theories are valuable if they are empirically accurate and possess some other cognitive values, whose definition is
traditionally always unclear.  If we adopt the complexity of the assumptions ${\cal C}$, as a cognitive value, then
two state of the art theories, say $T_1$ and $T_2$ are certainly valuable, and so are the predictions that
discriminate between them.  If the prediction of $T_1$ is confirmed, it makes the state of the art theory $T_2$
outdated, which produces progress according to Def.~\ref{def:SP}.  Hence, if we define valuable predictions as
those that discriminate between two state of the art theories, we capture the main motivation behind
\citep{Lakatos_1970}, while the progress that they signal also belongs entirely within the model $SP_0$.

Another classic goal of science is that of {\bf unification}.  But this also lacks a clear definition.  A unifying
theory $T_{\rm u}$ should at least reproduce all the phenomena described by the theories $T_1$ and $T_2$ that it
should unify.  But this is not a sufficient condition, because also the mere juxtaposition: $T_{\rm u}:= T_1 +
T_2$, fulfills it.  Characterizing a genuine synthesis is difficult, and the measure ${\cal C}$ is, as far as I
know, the only available well defined tool. If we adopt it, we can define a unification of $T_1$ and $T_2$ as a
theory $T_{\rm u}$ that describes at least all the phenomena described by $T_1$ and $T_2$, it is at least as
concise as $T_1 + T_2$, and it improves at least one of these aspects.  Finding such $T_{\rm u}$ is consistent with
the traditional idea of valuable unification and it also represents progress according to Def.~\ref{def:SP}.

Ad-hoc assumptions have always been the fatal problem in a rigorous application of the idea of {\bf
  falsifiability}.  Popper struggled unsuccessfully to find a semantic characterization of ad-hoc assumptions
\citep{Koertge1979}.  A characteristic feature of ad-hoc assumptions is that they are needed to explain only few
events, while valuable laws are used for a wide range of phenomena.  Those theories that try to describe more and
more data with more ad-hoc assumptions, and express them in ordinary languages\footnote{Because of the constraints
  of measurability, for realistic theories, that combine assumptions from many branches of science, it is
  practically impossible to check whether alternative formulations, radically different from the ordinary one,
  might enable much higher conciseness.}, necessarily grow their ${\cal C}(T)$ indefinitely.  On the other hand,
theories without ad-hoc assumptions can describe many data with no increase of ${\cal C}(T)$.  As a result, the
ratio: ${\cal C}(T)$ / (empirical accuracy of $T$) is typically much higher when $T$ contains many ad-hoc
assumptions.  Hence, the measure ${\cal C}(T)$ effectively penalizes ad-hocness, in a way that is consistent with
Popper's original motivation, and it offers a more solid justification to the idea and application of
falsifiability.

Another major goal of science is that of providing {\bf explanations}\footnote{There is actually a vast
  disagreement not only about the best general model of explanation \citep{sep-scientific-explanation}, but also
  whether many individual arguments are explanatory or not.  In fact, some authors want explanations to rely on
  causal sequences, which, in my opinion, assigns to the concept of {\em time} a fundamental philosophical role,
  that could conflict with some possible scientific theories; some authors want explanations to be easily
  understandable, which is hopelessly subjective; some authors want us to be able to add or suppress truly
  explanatory causes, which is an idea that ultimately relies on an understanding of {\em freedom}, which, however,
  systematically hides itself where science cannot go.  However, as far as the topic of this paper is concerned, I
  am not aware of any example of a theory $T$ that is better than $T'$, according to Def.~\ref{def:SP}, but less
  explanatory according to some other criteria.  On the other hand, it may well be that a theory $T$ is more
  explanatory than $T'$, in some sense, without any appreciable difference according to Def.~\ref{def:SP}.  This
  confirms that $SP_0$ may be incomplete but it only detects genuine progress (see below).}.  One of the dominant
views of explanation is the unificationist model \citep{Friedman1974-FRIEAS, Kitcher1989}.  In the version of
Friedman, the goal of science is to reduce the {\em number of independent laws} that we need to assume.  However,
it is always possible to collect any number of consistent laws into a single one (what I called the $\Xi$ trick).
In an attempt to solve this difficulty, Kitcher suggested that good explanations reduce the {\em number of
  different types of derivation patterns}.  But, it remains unclear how one should count the number of different
types of patterns: if we collect all the assumptions in a single one, we may claim that the needed patterns are
always just one or two.  The notion of complexity ${\cal C}$ proposed here is protected from these trivialization
arguments, because we are forced to use a syntax that refers to BMPs.  Under this constraint, it is hard to
formulate laws in a syntax that is radically different and more concise than the one commonly used in the best
scientific praxis.  As a result, the notion of complexity ${\cal C}$ captures the ideas of
\citet{Friedman1974-FRIEAS} and \citet{Kitcher1989} in a well defined way, and it produces a notion of progress
that is consistent with their spirit.

Another important view of progress is represented by {\bf truthlikeness} \citep{Oddie-SEP}, but an essential
requirement of truthlikeness is that scientific theories should be {\em bold}.  This is in fact the only reason why
truthlikeness is not expected to reward maximally a very accurate theory that consists in a collection of
experimental reports.  However, the concept of {\em boldness} lacks a definition, which corresponds to the
observation \citep{Oddie-SEP} that focusing only on the concepts of {\em content} and {\em likeness} is not enough.
The concept of conciseness proposed here is certainly very different from the concept of boldness, but when combined
with empirical accuracy, it is effectively able to penalize those very collections of reports, that boldness is
supposed to penalize.

\subsection{Other cognitive values?}
\label{sec:ocv}

Scientific progress, as defined in Def.~\ref{def:SP}, is truly unambiguous only to the extent that empirical
accuracy and the complexity ${\cal C}$ of the assumptions are the most relevant cognitive values; i.e., the other
values are either compatible or clearly less important.  In this section, I comment on those alternative cognitive
values, that are sometimes mentioned in the literature, and I argue that they are either not relevant for theory
selection and progress, or they are already taken into account by $SP_0$.

One cognitive value, that is often mentioned is that of {\em coherence} with other theories.  In this case we
should distinguish: {\bf logical coherence} cannot be matter of negotiation.  We are only considering theories that
are overall consistent (until evidence of the contrary)\footnote{The state of the art does contain different
  theories, that may be inconsistent with each other, but we never use them on the same deductive reasonings.}.
But, logical coherence is meaningful, strictly speaking, only within a single theory.  If we talk about logical
coherence of $T$ with $T'$, we are necessarily thinking about a larger (often more fundamental) theory that
encompasses both $T$ and $T'$.  In this sense, coherence is a first necessary step to unification, which is
consistent with $SP_0$, as discussed in the previous section.

A very different value is that of {\bf esthetic coherence}, which is closely related to the idea of {\bf symmetry}
and {\bf elegance}.  But --- to the extent that these notions are not totally subjective --- it is precisely when
the laws are more symmetric, or when the different parts of the body of science look more coherent, that it becomes
possible to achieve a simpler expression, in some language.  I am not aware of any case in which a scientist has
appealed to {\em elegance} or {\em beauty}, that could not be translated into {\em more conciseness}, in some
formulation.  A counterexample would be very interesting.
%% ch

Another value that is often mentioned is {\bf fruitfulness} \citep{KuhnET}.  But if we expect some ideas to lead,
eventually, to future successful theories, we should wait for this to happen before declaring progress!  It is
hardly appropriate, and certainly not necessary, to count as progressive those events that simply introduce some,
perhaps, promising ideas.  Even if --- in retrospect --- we must recognize those events as crucial for later
developments, it is often impossible to predict them in advance.  We expect a precise notion of progress to certify
what the scientists know to have achieved, not to anticipate what even the scientists themselves are still unable
to foresee.

Another aspect that is sometimes regarded as a sign of valuable scientific theories is {\bf a rich mathematical
  content}.  But, as far as I can tell, this can only refer to a system in which few assumptions allow the
derivation of many consequences.  But this is exactly what I have tried to define in this paper.

One might think that simple assumptions are very valuable in fundamental physics, but not necessarily so much in
other fields.  This impression arises because other fields also regard it as very valuable to use laws that may be
more complex themselves, but have a better {\bf fundamental justification}\footnote{Occasionally, these laws are
  also said to be more {\em ontologically plausible}, or more {\em natural}.  Unless these ideas refer to some {\em
    prejudices}, that cannot be considered valuable, they can only gain their justification from a more fundamental
  level.}.  But, having a fundamental justification means that these laws can be derived from more fundamental
ones, that need to be assumed anyway.  Hence, if I use laws with a fundamental justification, I need less higher
level assumptions, and I have reduced the amount of assumptions {\em overall}.  In conclusion, what seems to
contrast the tendency toward the simplification of the assumptions from a restricted perspective, it is actually
achieving greater simplification, when considering a broader scientific domain.

This does not mean that all disciplines should necessarily consider reductionism as their main goal.  I am claiming
here that the main goal is to reduce the assumptions.  The program of reduction of many phenomena, ultimately, to
particle physics has achieved unequaled successes, in reducing the overall assumptions in science.  But it does not
necessarily have to be the only way to progress, according to the view presented here.

In conclusion --- contrary to what is often claimed --- I see no evidence of a compelling need for other {\em
  independent} cognitive values, besides the two already included in the model $SP_0$.

\section{Concluding remarks}
\label{sec:conclusions}

In this paper I have argued that it is urgent and possible to identify the non-empirical cognitive values that
justify scientific theory selection.  I have also argued that the key is to understand the relation between
simplicity and measurability.

This realization has deep implications.  In particular it enables the formulation of a simple model for scientific
progress ($SP_0$) that is based only on empirical accuracy and conciseness.  This seems to agree with the spirit of
the other classic goals of science, and it has also enabled their precise definition.  To summarize the advantages
and limitations of $SP_0$ consider two questions.

First, do all cases of $SP_0$ represent truly relevant and progressive scientific results?  This obviously depends
on what we consider to be {\em relevant}, which is not only a scientific issue.  But, if we restrict our
consideration to phenomena that we assume to be relevant, then producing a new theory that describes the data more
accurately with less assumptions, and hence fulfills the conditions for progress in Def.~\ref{def:SP}, is certainly
a significant result.  In fact, we cannot satisfy the conditions of Def.~\ref{def:SP} by performing easy tasks,
such as trivial reformulations of old theories or the execution of unchallenging experiments.  Moreover, $SP_0$
measures unambiguous gain (with no loss) in cognitive values that are deemed necessary, according to the present
analysis.

Second, are there cases that we intuitively recognize as truly progressive, but are missed by $SP_0$?  I have
argued that most of the truly significant real episodes of progress are recognized as such, and I am not aware of
any {\em important} case that eludes $SP_0$, but I do not exclude that this may happen, as discussed in
Sec.~\ref{sec:limit}.

For these reasons, $SP_0$ might be perhaps {\em incomplete}, but it covers a very {\em significant} amount of
progressive cases\footnote{Note that previous precise characterizations could not go beyond a reductionist
  framework.}, since it was achieved quite often in the history of science.  Moreover, it only identifies {\em
  genuine} progress, since it cannot be fabricated easily.  As a result, it represents a challenging but realistic
goal for everyday research activity and can offer the philosophical ground for effective and general measures of
quality of scientific research.

This simple model may be surprising, as it is today widely accepted that the cognitive values cannot be reduced to
a few.  But the only support to this belief is the lack of a simpler, convincing account.  In order to refute the
model $SP_0$, one should provide evidence of at least one real case where $SP_0$ declares a theory $T$ better than
$T'$, but this conclusion is implausible, or a case where $SP_0$ suspends judgment although one theory should be
clearly better than the other.  Since I am not aware of any such evidence, this is a challenge to the reader.  
%% ch

Finally, I made no attempt to justify any cognitive value in terms of goals like {\em truth} or its approximations.
This is because I regard philosophy of science as a science itself, which sets to itself exactly the same goals
that it sets to science: to find the simplest principles able to describe as accurately as possible the empirical
data.  In the case of philosophy of science, the {\em data} are represented by our long and successful tradition of
scientific theories that were born, developed, provided great applications and were overthrown by better ones.
Coherently with this view, we have to identify those general values that have characterized our fruitful scientific
tradition along its whole history and across all disciplines.  The {\em values} are not judged a priori, they are
judged by their historical fruits\footnote{Of course, one can still claim that this is all consistent with
  \citet{Kuhn1}, because the values of {\em empirical accuracy} and {\em simplicity of the assumptions} define,
  after all, just another (big) paradigm, and it is only within this paradigm that theories can be evaluated.  I
  have no objection, as long as we recognize that this paradigm {\em embraces all of science}.}.

\appendix
\section{Empirical comparisons}
\label{sec:app}

For completeness, this appendix defines the relative empirical accuracy of two scientific theories, although
the intuitive meaning should be sufficient, for the purposes of this paper.

In science, we often want to compare theories that are not logically equivalent and may even lack any common
logical substructure.  However, to be comparable, they need to deal with some common topics, that is, they need at
least to share some set ${\cal E}$ of MPs.  More precisely, there should be a one-to-one correspondence ${\cal J}$
between ${\cal E}$ of $T$ and ${\cal E}'$ of $T'$, which are also interpreted in the same way (i.e., their
respective extensions should coincide)\footnote{Some MPs in one theory may have no correspondence in the other.
  But, either we can reduce the comparison to different laws concerning common MPs, or we must conclude that the
  two theories deal with different topics.  For example, {\em absolute time} coordinates are meaningful for
  Galilean relativity and they are not for special relativity.  But we can compare the different laws of the two
  theories in terms of {\em space-time} coordinates, that are meaningful for both.}.

\begin{defn}
\label{def:EA}
\emph{(Comparability and relative empirical accuracy).}  We say that $T$ and $T'$ are {\em comparable} relatively
to the MPs ${\cal E}$ of $T$ and ${\cal E}'$ of $T'$ if there is a one-to-one correspondence ${\cal J}$ between
${\cal E}$ and ${\cal E}'$ and the experimental outcome relative to ${\cal E}'={\cal J}({\cal E})$ are interpreted
in the same way by $T$ and $T'$.  We say that $T$ is {\em empirically more accurate} than $T'$, with respect to the
MPs ${\cal E}$, if $T$ and $T'$ are comparable, and if all the results of $T$ concerning ${\cal E}$ match the
experiments at least as well as those of $T'$ concerning ${\cal E}'$, but they match them better at least in one
case.
\end{defn}

\begin{acknowledgements}
I am very grateful to all the people that shared their comments on older drafts of this paper and on my
presentations.  In particular, I am especially indebted to Gustavo Cevolani, Vincenzo Crupi and Georg Schiemer.
\end{acknowledgements}

\bibliography{../../philo}{}
\bibliographystyle{chicago}

\end{document}